\documentclass[prl,twocolumn,twoside,showpacs]{revtex4-1}

\usepackage{hyperref}
\usepackage{amsmath}
\usepackage{graphicx}
\usepackage{color}

\usepackage[varg]{txfonts}

\begin{document}

\title{Microscopic calculation of the $^3$He($\alpha$,$\gamma$)$^7$Be and $^3$H($\alpha$,$\gamma$)$^7$Li capture cross sections\\ using realistic interactions}

\author{Thomas Neff}
\email[Electronic address: ]{t.neff@gsi.de}
\affiliation{GSI Helmholtzzentrum f{\"u}r Schwerionenforschung GmbH,
  Planckstra{\ss}e 1, 64291 Darmstadt, Germany}

\date{\today}

\renewcommand{\vec}[1]{\mathbf{#1}}

\newcommand{\op}[1]{#1}
\newcommand{\ket}[1]{\big| {#1} \big> }
\newcommand{\ketintr}[1]{\ket{#1}}
\newcommand{\braket}[2]{\big< {#1} \big| {#2} \big> }

\newcommand{\fm}{\ensuremath{\text{fm}}}
\newcommand{\keV}{\ensuremath{\text{keV}}}
\newcommand{\MeV}{\ensuremath{\text{MeV}}}
\newcommand{\nuc}[2]{\ensuremath{{}^{#1}\textrm{#2}}}

\newcommand{\Healphagamma}{$^3$He($\alpha$,$\gamma$)$^7$Be}
\newcommand{\Halphagamma}{$^3$H($\alpha$,$\gamma$)$^7$Li}

\begin{abstract}
The radiative capture cross sections for the $^3$He($\alpha$,$\gamma$)$^7$Be and $^3$H($\alpha$,$\gamma$)$^7$Li reactions are calculated in the fully microscopic fermionic molecular dynamics approach using a realistic effective interaction that reproduces the nucleon-nucleon scattering data. At large distances bound and scattering states are described by antisymmetrized products of $^4$He and $^3$He/$^3$H ground states. At short distances the many-body Hilbert space is extended with additional many-body wave functions needed to represent polarized clusters and shell-model-like configurations. Properties of the bound states are described well, as are the scattering phase shifts. The calculated $S$ factor for the $^3$He($\alpha$,$\gamma$)$^7$Be reaction agrees very well with recent experimental data in both absolute normalization and energy dependence. In the case of the $^3$H($\alpha$,$\gamma$)$^7$Li reaction the calculated $S$-factor is larger than available experimental data by about 15\%.
\end{abstract}

\pacs{25.55.-e,21.60.De,27.20.+n,26.20.Cd}

\maketitle


The \Healphagamma\/ reaction is one of the key reactions in the solar proton-proton chains \cite{adelberger98,adelberger10}. It competes with the \nuc{3}{He}(\nuc{3}{He},2$p$)\nuc{4}{He} reaction and therefore determines the production of \nuc{7}{Be} and \nuc{8}{B} neutrinos in the ppII and ppIII branches. For a long time the experimental situation regarding the capture cross section was not clear due to conflicting experimental results \cite{adelberger98}. In recent years the capture cross section has been remeasured at the Weizmann institute \cite{narasingh04}, by the LUNA Collaboration \cite{bemmerer06,confortola07}, by the Seattle group \cite{brown07}, and by the ERNA Collaboration \cite{dileva09} now providing consistent high precision data. Nevertheless, it is still not possible to reach the low energies relevant in solar burning and the data have to be extrapolated with the help of models. A careful analysis of the new data sets and a discussion of the extrapolation and its uncertainties is given in Ref.~\cite{adelberger10}.

The first attempts to model the capture cross sections were done by using an external capture model \cite{christy61,tombrello63a} where only the asymptotic form of the bound and scattering state wave functions enters, neglecting the behavior of the wave function at short distances. In potential models like, e.g., Ref.~\cite{kim81} the wave functions are described by two pointlike clusters interacting via an effective nucleus-nucleus potential which is adjusted to give reasonable properties for the bound states and the scattering phase shifts. In the framework of the microscopic cluster model, e.g., Refs.~\cite{liu81,langanke86,mertelmeier86,kajino86}, the system is described by antisymmetrized wave functions of two clusters. One has to solve for the relative motion of the clusters by using resonating group or generator coordinate methods. In these microscopic models phenomenological nucleon-nucleon interactions are used. Like in the potential models these interactions are tuned to reproduce certain properties of bound and scattering states within the restricted cluster model space. There have been attempts \cite{mertelmeier86,csoto00} to go beyond the single-channel approximation by including the \nuc{6}{Li}+$p$ channel, but such enlarged model spaces require again modifications of the phenomenological interaction.

Predictive power is expected from \emph{ab initio} methods which use realistic interactions that reproduce the nucleon-nucleon scattering data and the deuteron properties. Solving the many-body problem with realistic interactions is hard, as very large model spaces are required and up to now consistent \emph{ab initio} reaction calculations have been possible only for single nucleon projectiles \cite{nollett07,quaglioni09}. The \Healphagamma\/ reaction was studied in hybrid approaches, where asymptotic normalization coefficients calculated from \nuc{7}{Be} bound state wave functions using variational Monte Carlo \cite{nollett01} and the no-core shell-model \cite{navratil07b} were combined with conventional potential models. None of these calculations is successful in describing both the normalization and the energy dependence of the capture cross section data.

In this Letter, we present the first \emph{ab initio} type calculation of the \Healphagamma\/ and \Halphagamma\/ capture cross sections. We describe consistently bound and scattering states starting from a realistic effective interaction derived in the unitary correlation operator method. The fermionic molecular dynamics approach is used to create many-body wave functions that capture the relevant physics in the interaction region. Frozen cluster configurations with \nuc{4}{He} and \nuc{3}{He}/\nuc{3}{H} ground states are used at large distances.


The effective interaction is derived from the realistic Argonne~V18 interaction \cite{wiringa95}, that reproduces the deuteron properties and the nucleon-nucleon scattering phase shifts. The interaction is transformed into a phase-shift equivalent low-momentum interaction by using the unitary correlation operator method (UCOM) \cite{ucom98,*ucom03,ucom10} where short-range central and tensor correlations are incorporated explicitly. In this work we use UCOM correlation functions that are derived from a Hamiltonian evolved using the similarity renormalization group (SRG) as described in Ref.~\cite{ucom10} with a flow parameter $\alpha\!=\!0.20\,\fm^4$, corresponding to a soft cut-off $\lambda\!=\!1.5\,\fm^{-1}$. As shown in Ref.~\cite{ucom10}, no-core shell-model calculations using the two-body UCOM(SRG) interaction are able to reproduce the binding energies of triton, \nuc{4}{He} and \nuc{7}{Li}.


Fermionic Molecular Dynamics (FMD) is a microscopic many-body approach that has been used successfully for nuclear structure studies of nuclei in the $p$ and $sd$ shell. See \cite{hoyle07,geithner08,zakova10} for some recent applications and \cite{fmd08} for a general discussion. FMD is based on intrinsic many-body basis states that are Slater determinants
\begin{equation}
  \label{eq:fmdslaterdet}
  \ket{Q} = \op{\mathcal{A}} \bigl\{ \ket{q_1} \otimes \ldots \otimes
    \ket{q_A} \bigr\} \: ,
\end{equation}
with Gaussian wave packets as single-particle states
\begin{equation}
  \braket{\vec{x}}{q_k} = 
  \exp \biggl\{ -\frac{(\vec{x} -\vec{b}_k)^2}{2 a_k} \biggr\} \otimes
  \ket{\chi^\uparrow_{k},\chi^\downarrow_{k}} \otimes \ket{\xi_k} \: .
\end{equation}
The complex parameters $\vec{b}_k$ encode the mean positions and momenta of the wave packets. The width parameters $a_k$ are variational and can be different for each nucleon. The spin can assume any direction, and isospin $\xi_k$ is $\pm 1/2$. The wave packet basis is very flexible and contains harmonic oscillator shell-model and Brink-type cluster states as special cases.

To restore the symmetries of the Hamiltonian the intrinsic basis states $\ket{Q}$ are projected on parity, angular momentum and total linear momentum 
\begin{equation}
  \ketintr{Q; J^\pi MK} \otimes \ket{\vec{P}_\mathrm{cm}=0} =
  \op{P}^{J}_{MK} \op{P}^\pi \op{P}^{\vec{P}=0}\ket{Q} \: ,
\end{equation}
so that the wave function factorizes into the internal part and the center-of-mass motion given by a plane wave.

In general the intrinsic states have no axial symmetry and $K$ is not a good quantum number. Linear dependent combinations among the different $K$-projections have to be removed. This is done numerically and introduces a small ambiguity in the size of the model space. We will exploit this ambiguity later to fine-tune the \nuc{7}{Be} and \nuc{7}{Li} binding energies.

All bound and scattering states are represented by using a set of intrinsic states
$\ket{Q^{(i)}}$
\begin{equation}
  \ket{\Psi; J^\pi M \alpha} =
	\sum_{iK} \ketintr{Q^{(i)}; J^\pi MK} \: C^{J^\pi \alpha}_{iK} \: .
\end{equation}
Proper boundary conditions for bound and scattering states are imposed by using the microscopic $R$-matrix approach developed by the Brussels group \cite{baye77,*baye83,descouvemont10}.

\begin{figure}
  \includegraphics[width=0.95\columnwidth]{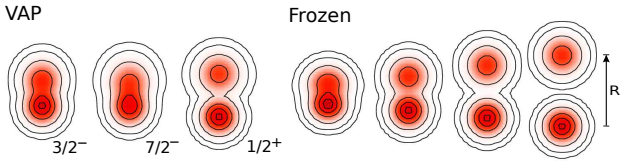}
  \caption{(color online). Cuts through the density distributions of intrinsic basis states. Left: Selected polarized configurations obtained in variation after angular momentum and parity projection for $3/2^-$, $7/2^-$ and $1/2^-$ states. Right: Frozen configurations where only the cluster distance is varied.}
  \label{fig:intrinsic}
\end{figure}

At large distances the \nuc{7}{Be} and \nuc{7}{Li} wave functions consist of \nuc{4}{He} and \nuc{3}{He}/\nuc{3}{H} clusters in their ground states interacting via the Coulomb interaction only. The relative motion of these frozen clusters is therefore given by Whittaker and Coulomb functions for bound and scattering states, respectively. Microscopically we describe these cluster configurations with FMD Slater determinants where the clusters are put at a distance $R$. The wave functions of the individual clusters are obtained by variation in the FMD model space.

In the interaction region the nuclear interaction will polarize the clusters. To include these polarization effects we extend the model space with additional FMD basis states obtained by variation after projection (VAP) on spin-parity $3/2^-$, $1/2^-$, $7/2^-$, and $5/2^-$ as well as on $1/2^+$, $3/2^+$, and $5/2^+$. The square radius of the intrinsic state is used as a constraint to generate configurations corresponding to cluster distances from 1 to 5~fm. Together with the frozen configurations that extend to distances slightly beyond the channel radius $a\!=\!12\,\fm$, we have about 50 intrinsic basis states to represent the inner part of the wave function. Density distributions of typical frozen and polarized basis states are shown in Fig.~\ref{fig:intrinsic}.


\begin{table}[b]
\caption{Calculated and experimental bound state properties. Energies with respect to the \nuc{4}{He}-\nuc{3}{He} and \nuc{4}{He}-\nuc{3}{H} thresholds, respectively. Experimental charge radii are from Refs.~\cite{noertershaeuser09,sanchez06}, the \nuc{7}{Li} quadrupole moment from Ref.~\cite{diercksen88}.}
\begin{ruledtabular}
\begin{tabular}{c|cc|cc}
& \multicolumn{2}{c|}{\nuc{7}{Be}} & \multicolumn{2}{c}{\nuc{7}{Li}} \\
& FMD & Exp & FMD & Exp\\
\hline
$E_{3/2^-}$ [\MeV] & -1.49 & -1.586 & -2.39 & -2.467 \\
$E_{1/2^-}$ [\MeV] & -1.31 & -1.157 & -2.17 & -1.989 \\
$r_\mathrm{ch}$ [\fm] & 2.67 & 2.647(17) & 2.46 & 2.390(30) \\
$Q$ [$e\,\fm^2$] & -6.83 & & -3.91 & -4.00(3) \\
\end{tabular}
\end{ruledtabular}
\label{tab:observables}
\end{table}

When the model space is restricted to frozen configurations the $3/2^-$ and $1/2^-$ states in \nuc{7}{Be} are bound by only 240 and 10~keV respectively. The FMD VAP configurations are therefore essential to a get a good description of the bound states. As mentioned the numerical elimination of linear dependent states in the $K$-mixing procedure introduces a small ambiguity in the model space size that translates into an ambiguity in the binding energy of about 150~keV. As the reaction cross section depends very sensitively on phase space we exploit this ambiguity to tune the centroid of the $3/2^-$ and $1/2^-$ bound state energies to the experimental value. The calculated splitting between the bound states is too small compared to the data. However, the total cross section essentially depends only on the centroid energy, whereas the branching ratio slightly changes with the splitting. The bound state properties for \nuc{7}{Be} and \nuc{7}{Li} are summarized in Table~\ref{tab:observables}. The charge radii and quadrupole moments test the tail of the wave functions and agree reasonably well with experiment.

\begin{figure}[t]
  \includegraphics[width=0.95\columnwidth]{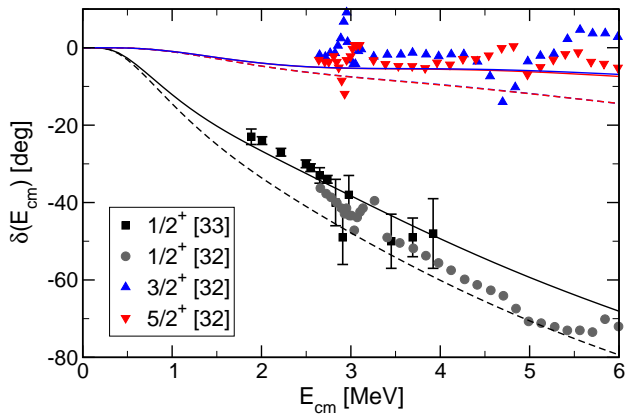}
  \caption{(Color online). \nuc{4}{He}-\nuc{3}{He} scattering phase shifts. Dashed lines show results using only frozen configurations, solid lines are obtained with the full FMD model space. The calculated $D$-wave phase shifts lie on top of each other. Experimental results are from Refs.~\cite{spiger67,boykin72}.}
  \label{fig:phaseshiftshe3alpha}
\end{figure}

In Fig.~\ref{fig:phaseshiftshe3alpha}, we show the phase shifts for scattering in the $S$- and $D$-wave channels. As for the bound states, the addition of polarized configurations to the model space significantly changes the results and leads to a good agreement with the available data \cite{spiger67,boykin72}.

\begin{figure}[b]
  \includegraphics[width=0.95\columnwidth]{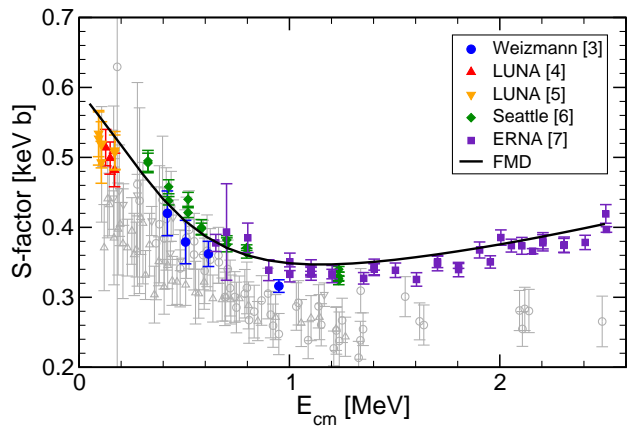}
  \caption{(Color online). The astrophysical $S$ factor for the \Healphagamma\/ reaction. The FMD result is given by the solid line. Recent experimental data \cite{narasingh04,bemmerer06,confortola07,brown07,dileva09} are shown as dark colored symbols and older data \cite{adelberger98} as light symbols.}
  \label{fig:sfactorhe3alpha}
\end{figure}

The capture cross section for the \Healphagamma\/ reaction is calculated by using the many-body scattering and bound eigenstates of the Hamiltonian. In the energy range up to 2.5~MeV, it has been shown \cite{nollett01} that only dipole transitions from the $S$- and $D$-wave scattering states have to be considered. The obtained $S$ factor is shown in Fig.~\ref{fig:sfactorhe3alpha} together with the experimental data. Our results are in good agreement with the recent measurements regarding both the absolute normalization and the energy dependence. The extrapolated zero-energy $S$ factor is $S_{34}(0)\!=\!0.593\,\keV\,\mathrm{b}$.

\begin{figure}[b]
  \includegraphics[width=0.95\columnwidth]{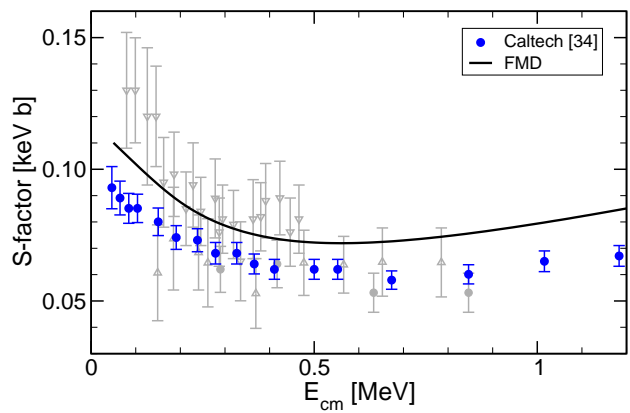}
  \caption{(Color online). The astrophysical $S$ factor for the \Halphagamma\/ reaction. The FMD result is given by the solid line. Most recent experimental data are shown as dark symbols and older data as light symbols (\cite{brune94} and references therein).}
  \label{fig:sfactorh3alpha}
\end{figure}

As our model successfully describes the \Healphagamma\/ reaction, it should also do well for the isospin mirror reaction \Halphagamma\/. As shown in Fig.~\ref{fig:sfactorh3alpha} we observe a good agreement for the energy dependence of the $S$-factor but find that the absolute normalization is about 15\% larger than 
the data by Brune, Kavanagh, and Rolfs \cite{brune94}.

In summary our calculations are able to describe consistently the bound state properties, and the scattering phase shifts as well as the normalization and energy dependence of the \Healphagamma\/ capture cross section. Our results deviate from the correlation between the ground state quadrupole moment and zero-energy $S$ factor found in cluster models using phenomenological interactions \cite{kajino86,csoto00}. Our approach differs in two main aspects from those earlier studies. First, we use a well defined effective interaction that describes the nucleon-nucleon scattering data. In contrast to phenomenological effective interactions the UCOM interaction has a pronounced momentum dependence and a longer range due to the explicitly included pion exchange, a feature that turns out to be important for the low energy scattering solutions. Second, our model space is larger than in the cluster model. Additional FMD basis states in the interaction region describe polarized clusters and shell-model-like configurations. Although they are only a small admixture in the full wave functions they are essential to describe the bound state properties as well as the scattering phase shifts.

\begin{figure}
  \includegraphics[width=0.95\columnwidth]{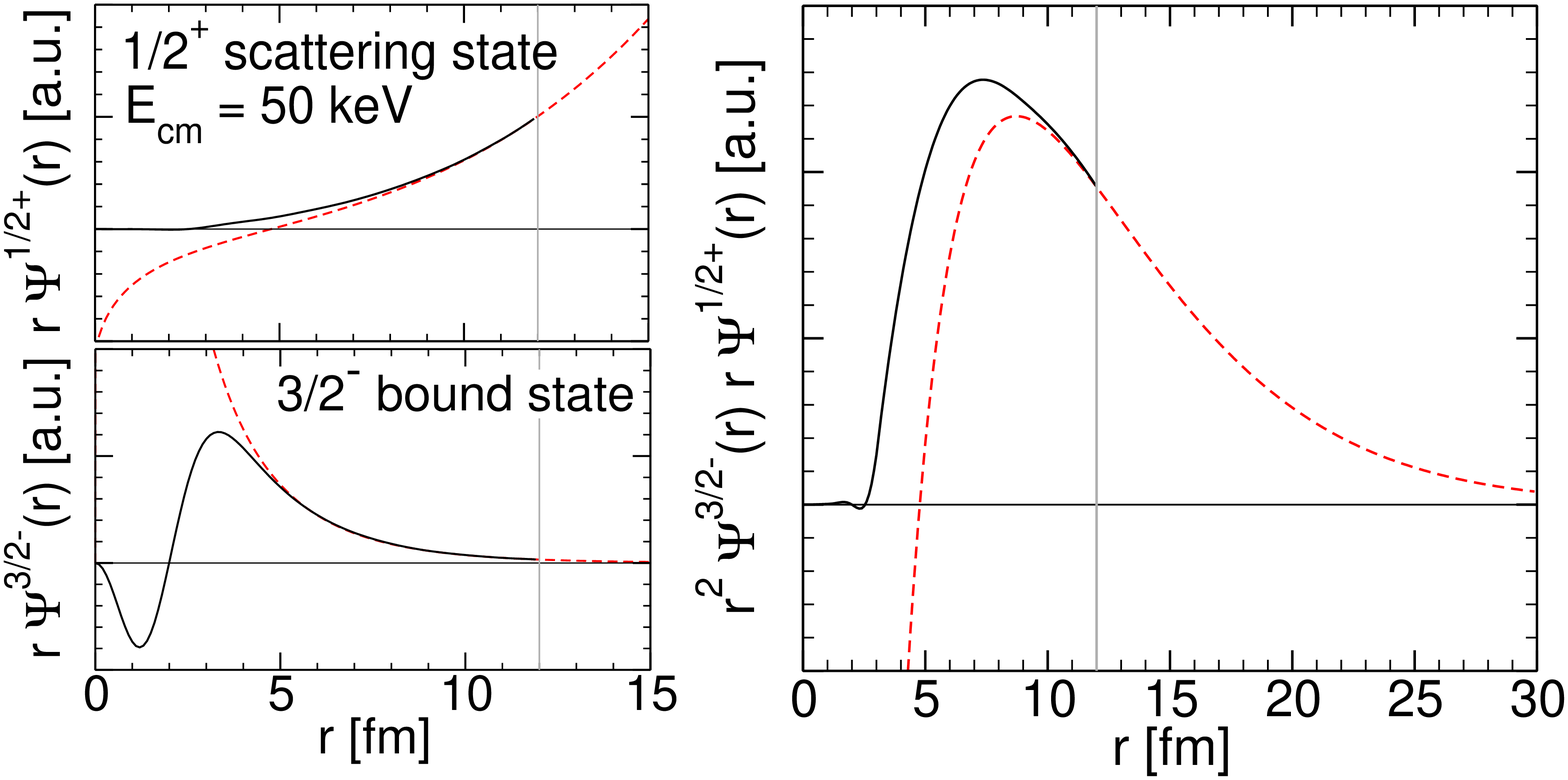}
  \caption{(Color online). On the left: \nuc{7}{Be} overlap functions for a low energy $1/2^+$ scattering state and the $3/2^-$ bound state (solid lines). Coulomb and Whittaker functions matched at the channel radius (dashed lines). On the right: Dipole strength calculated with overlap functions (solid line) and with Coulomb and Whittaker functions (dashed line).}
  \label{fig:dipolemes}
\end{figure}

The results can also be studied in terms of overlap functions that are obtained by mapping the microscopic many-body wave functions onto the relative wave function of two pointlike nuclei in the resonating group formalism. In Fig.~\ref{fig:dipolemes} we show the overlap functions for the $1/2^+$ scattering state at $E_\mathrm{cm}\!=\!50\,\keV$ and the $3/2^-$ bound state. The nodes in the overlap functions reflect the antisymmetrization between the clusters. We also show the dipole strength calculated with these overlap functions. It reproduces the dipole matrix element calculated with the microscopic wave functions within 2\%. Comparing with the dipole strength obtained from the Coulomb and Whittaker functions matched at the channel radius we observe sizable differences up to distances of about 9~fm. This indicates that the assumption of predominant external capture at low energies is not that well satisfied.

Future calculations should investigate the role of three-body forces. It is expected that low-momentum three body forces would increase the splitting between the $3/2^-$ and $1/2^-$ states but would have a minor effect on the centroid energy. Furthermore more detailed wave functions could be used. In the FMD approach it is difficult to describe long-range tensor correlations explicitly, so that the absolute binding energies are underestimated, although the binding energy with respect to the cluster threshold is in very good agreement with no-core shell-model results. Nevertheless, we expect that such improvements will not change the capture cross sections significantly as important properties like phase shifts of the scattering states, binding energy with respect to the cluster threshold, asymptotic behavior of the bound state wave functions as tested by charge radius and quadrupole moment, and proper treatment of antisymmetrization are already well described in the present calculation.


\end{document}